\documentclass[a4paper]{article}
\usepackage[T2A]{fontenc}
\usepackage[cp1251]{inputenc}
\usepackage{cite,enumerate,float,indentfirst}
\usepackage{graphicx}
\usepackage{booktabs, tabularx}
\usepackage{mathtools}
\usepackage{amssymb,amsfonts,amsmath,mathtext}

\begin{document}
\title{An evolutionary game with environmental feedback and players' opinions}
\author{Lorits E. M., Gubar E. A.}
\date{}
\maketitle

\vskip -60pt

\section{Introduction}
Evolutionary games are a developing subfield of game theory~\cite{Cress, Brown}. Evolutionary games are used to model change in large but finite populations, where all agents have biological, social or economic characteristics that determine their behaviour. Furthermore, each agent is assumed to have no significant effect on the state of the population.

Evolutionary game theory is widely used in many scientific fields. For example, a book~\cite{Broom} was published in 2022 that collected many evolutionary models of real biological processes. In addition, evolutionary games
are used to simulate the interaction of large numbers of agents in a network~\cite{wirelessNetw, krivan,set, Ming}. In medicine, evolutionary games can be used, for example, to find methods to fight cancer~\cite{cancer1, cancer2, cancer3} or to solve the problem of vaccinating the population~\cite{matmod, Gubar}.

Papers~\cite{E7518, 1803, krzys} consider the population and the changes that occur in that population, taking into account the environment and the state of that environment. In addition, the effect of environmental feedback on a population has been studied extensively~\cite{Zhiliang,Meng,spb}. The paper explores the idea of how the state of the environment and agents' opinions about it affect the state of the system. The population, the environment and agents' opinions form a hierarchical structure, where a change in one parameter of the system responsible for the state of the environment, the population or agents' opinions, causes a change in the remaining elements of the system. On the one hand, the state of the environment depends on the prevalence of a particular type of behaviour in the population. On the other hand, the state of the population depends on the popularity of the opinions in the population. The popularity of agents' opinions depends on the state of the environment and the population. In this paper, we consider the state of the environment and the popularity of the opinions as control parameters on the population dynamics.

\section{An evolutionary game}
Consider a population of size $N$ existing in a finite space. It is assumed that the state of the population changes as a result of random pairwise interactions between its agents. It is also assumed that the number of agents is large and that each individual agent has no significant effect on the population~\cite{Weibull, pged}. Another assumption is that the population has two types of behaviour that agents can follow. An agent's choice of the $i$th type of behaviour is similar to the choice of the $i$th pure strategy in a non-cooperative game. It leads to a partition of the population into two subgroups. The agents of each subgroup are programmed to use the same pure strategy. The state of the population is defined as a vector $x_N (t) = (x_1 (t), x_2 (t))$, where each component $x_i(t)$ is the fraction of the population using the pure strategy $i$. This vector can be thought of as a mixed population strategy~\cite{Weibull}. Let $x(t) = x_1(t)$, then $x_2(t) = 1 - x(t)$. The payoffs of the agents refer to the number of offspring (in biological systems) or the number of successors (in economic and social systems) that follow the net strategy $i$.
Over time, random pairwise meetings between agents occur in the population.

The outcomes of these encounters can be described by a bimatrix game~\cite{Petrosyan}. Traditionally, in evolutionary games, it is customary to consider all processes on behalf of the first player, so everything below is formulated in terms of the first player. Let $e^i$ be the vector corresponding to the $i$th net strategy of the player. The $i$th element is one and all others are zero. We introduce the function $u(e^i,x_N) = e^i \cdot Ax, i=1,\ldots, n$ as the expected payoff of the agent with pure strategy $i$ when facing a random opponent. This payoff depends on the population state vector $x_N$. Based on the payoff of the randomly chosen agent, the corresponding average payoff of the population is determined
\begin{equation}
u(x_N,x_N) = \sum\limits_{i\in K} x_iu(e^i,x_N), i=1,\ldots, n.
\end{equation}

The change in population composition corresponds to a change in the proportion of agents adhering to net strategy $i$. These changes are described by the replicator dynamics equation~(\ref{x}), depending on the fractional distribution of players in the population $x_N$ and the first player's payoff matrix $A$.
\begin{equation}
 \dot x =x(1-x)(u(e^1,x_N) - u(e^2,x_N)).
 \label{x}
\end{equation}

In this paper, the population is assumed to be dependent on environmental feedbacks that affect the expected payoffs of agents. The resources available to the agents are considered as the environment. The state of the environment is described by the parameter $n(t)$, $n\in [0, 1]$, where $n = 0$ $(n = 1)$ when the environment is completely consumed (replenished). The change in the state of the environment is determined by the dynamics~(\ref{n1}) proposed in~\cite{1803}. It depends on the change in the state of the population.
\begin{equation}
 \dot n = n(1-n)(\theta x + \psi (1-x)),
 \label{n1}
\end{equation}
where the multiplier $n(n - 1)$ corresponds to the logistic growth of the environmental parameter. Depending on the state of the population $x_N$, resources are replenished or depleted. Agents following the first pure strategy replenish resources n at a rate $\theta > 0$, while agents following the second pure strategy cause resources to be depleted at a rate $\psi = -1$.

The player payoff matrix $A_n$ establishes the relationship between the population and the state of the environment~\cite{E7518}.
\begin{eqnarray}
A_n = 
  \begin{pmatrix}
    a_{11}^n & a_{12}^n\\
    a_{21}^n & a_{22}^n
  \end{pmatrix}
  = nA_1 + (1-n)A_0 
  = n
    \begin{pmatrix}
    a_{11}^1 & a_{12}^1\\
    a_{21}^1 & a_{22}^1
  \end{pmatrix}
  + (1-n)
    \begin{pmatrix}
    a_{11}^0 & a_{12}^0\\
    a_{21}^0 &a_{22}^0
  \end{pmatrix}.
\label{An}
\end{eqnarray}

When $n = 1$ ($n = 0$), the game is defined by the payoff matrix $A_1$ ($A_0$). The matrices $A_1$ and $A_0$ are set so that the non-cooperative game, which takes place between agents at random encounters in the population, has different Nash equilibrium positions in replenished and depleted environments.

An article~\cite{1803} examines changes in population state using replicator dynamics, which takes into account feedback from the environment and depends on public opinion. Opinion reflects population agents' awareness of the state of the environment.
In contrast to this study, the current work assumes that each agent has its own personal opinion about the state of the environment, but does not have reliable information about it.

Consider the case where an agent can hold one of two opinions, $m_1$ or $m_2$, regardless of the strategy it chooses. We define the distribution of opinions in the population as a vector $y_N (t) = (y_1 (t), y_2 (t))$, where each component $y_i (t)$ is the proportion of agents in the population holding opinion $m_i$. For convenience, we denote $y_1 = y(t)$, $y_2 (t) = 1 - y(t)$. The process of opinion distribution in a population can be described by mean dynamics, which allows changes in the population to be described by a proportional imitation rule~\cite{pged}.

Since any imitation protocol is subject to noise, i.e. errors in estimating the opponent's expected payoff, it is assumed that an agent can have different levels of confidence in the opinions of opponents, depending on the opponent's strategy.

Let's introduce a matrix $B$, whose elements $b_{ij}$ represent the degree of confidence of the agent with opinion $m_i$ in the agent pursuing strategy $j$. Based on a pairwise imitation protocol~\cite{Ming}, an imitation protocol~(\ref{p}) has been designed to describe changes of opinion popularity in the population.

Let us determine the expected playoff of the agent with opinion $m_i$ to determine the pairwise imitation protocol.

Since an agent with opinion $m_i$ can follow either the first or the second type of behaviour in the population, its average payoff is obtained by multiplying the proportions of agents using its strategy $j$ by the expected payoff of the player with the corresponding type of behaviour: $x_ju(e^j,x_N,A_n)$ given the coefficients of the confidence matrix $B$. Thus, the average payoff of the agent with opinion $m_i$ is given by 
\vskip-10pt
\begin{equation}
\sum\limits_{j=1}^2 x_ju(e^j,x_N,A_n)b_{ij}.
\end{equation}

Thus, the expected playoff of an agent with an opinion of $m_i$ can be written as
\begin{equation}
y_i\sum\limits_{j=1}^2 x_ju(e^j,x_N,A_n)b_{ij}.
 \label{playoffY}
\end{equation}

Consequently, the imitation protocol is:
\begin{equation}
p_{ij} = \left[y_j\sum\limits_{l=1}^2 x_lu(e^l,x_N,A_n)b_{jl} - y_i\sum\limits_{l=1}^2 x_lu(e^l,x_N,A_n)b_{il}\right]_0^1.
 \label{p}
\end{equation}

Thus, we get the dynamics, which describes the popularity of the opinions in the population based on the mean dynamics~(\ref{y}) and pairwise imitation protocol:
\begin{equation}
 \dot y =(1-y)p_{21}-yp_{12}.
 \label{y}
\end{equation}

It is assumed that the state of the population depends on the popularity of opinions. Under this assumption, the first player's payoff matrix can be rewritten in the form $A_y = yA_1 + (1-y)A_0$, which is obtained from the matrix~(\ref{An}) if the parameter $n$ representing the state of the environment is replaced by the fraction of players $y$ with an opinion $m_1$.

Thus, an evolutionary game with environment-opinion feedback can be represented as
\begin{equation}
\left\{
\begin{aligned}
    \dot x &=x(1-x)(u(e^1,x_N,A_y) - u(e^2,x_N,A_y)),\\
    \dot n &=n(1-n)(\theta x + \psi (1-x)),\\
    \dot y &=(1-y)p_{21}-yp_{12}.
\end{aligned}\right.
\label{syst}
\end{equation}

\section{Example 1. Hawk-Dove game}

As mentioned in the previous chapter, random pairwise encounters of population agents are described by a bimatrix game of two individuals. In the numerical experiments, bimatrix games with a known structure were considered. For these games, Nash equilibrium positions are obtained theoretically. This allows us to estimate the effect of environmental feedback and the opinions of the agents on the population.

In the current experiment, the Hawk-Dove game is chosen as the base game describing the interaction of the agents. The parameter $v$~-- is the value of the resource, the parameter $c$~-- is the cost of resources.

Since the model takes into account changes in the population state as a function of agents' opinions, according to the formula~(\ref{An}), we need to introduce two payoff matrices

\begin{equation}
A_0 =
  \begin{pmatrix}
 \frac{v_0-c_0}{2} & v_0\\ 0& \frac{v_0}{2}
  \end{pmatrix},
  A_1 =
  \begin{pmatrix}
\frac{v_1-c_1}{2} & v_1\\  0& \frac{v_1}{2}
  \end{pmatrix}.
  \label{A0A1HD}
\end{equation}

The Hawk-Dove game is characterized by three Nash equilibria: two asymmetric equilibria in the pure strategies $(e^1,e^2)$, $(e^2,e^1)$ and one symmetric equilibrium in the mixed strategies $(x_N,x_N)$, where $x_N=(\frac{v}{c}, 1 - \frac{v}{c})$~\cite{matmod}. In this case the average population payoff is $u(x_N,x_N) = \frac{v}{2}-\frac{v^2}{2c}$.

In the numerical experiment, the following parameters are used for the system~(\ref{syst}): 
\vskip-20pt
\begin{table}[H]
\caption{System parameters.\label{tab1}}
\newcolumntype{C}{>{\hsize=.27\hsize\centering\arraybackslash}X}
\newcolumntype{K}{>{\arraybackslash}X}
\small
\begin{tabularx}{\textwidth}{CCK}
\toprule
\textbf{Parameter}	& \textbf{Value}  & \textbf{Description}\\
\midrule
$\theta$   &2 & The rate of replenishment of resources by hawks\\
$\psi$ & -1 & The rate at which doves consume resources\\
$x_0$   &0.5 &Proportion of hawks in the population at the initial time moment\\
$n_0$   &0.3  & Environmental condition at initial time moment \\
$v_0$ &4 & The value of the resource\\
$c_0$ &12 &The cost of resources\\
$v_1$ &7 &The value of the resource\\
$c_1$ &10 &The cost of resources\\
$B$ & $\left( \begin{smallmatrix}    0.5 & 0\\ 0 & 0.5\end{smallmatrix} \right)$ &  Agents with opinion $m_1$ ($m_2$) are assumed to trust only hawks (doves)\\
\bottomrule
\end{tabularx}
\end{table}
\unskip

\begin{figure}[h!]
\begin{minipage}[h]{0.49\linewidth}
\center{\includegraphics[width=1\linewidth]{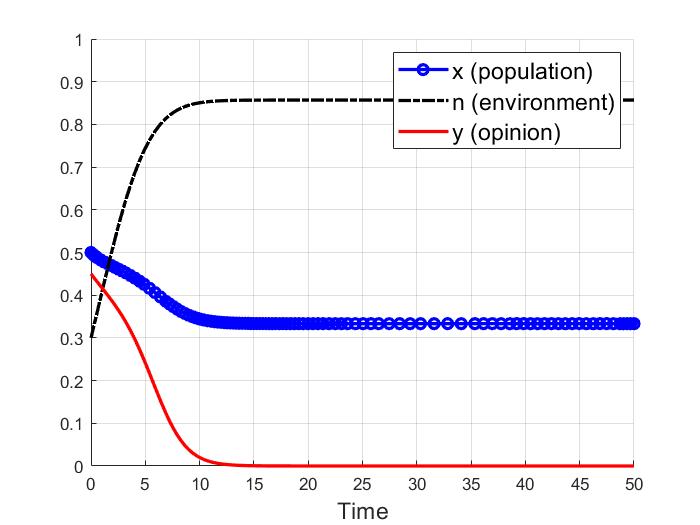}} а) $y(0) = 0.45$ \\
\end{minipage}
\hfill
\begin{minipage}[h]{0.49\linewidth}
\center{\includegraphics[width=1\linewidth]{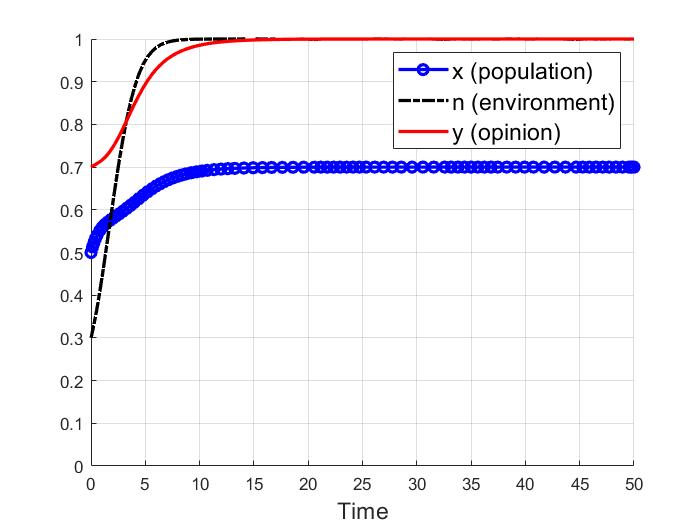}} б) $y(0) = 0.7$\\
\end{minipage}
\caption{The dependence of the state of the system on the initial value of the part of the population that holds the view $m_1$ $y(0)$; $x(0) = 0.5$, $n(0)=0.3$.}
\label{ris2:0503yB2}
\end{figure}

Figure~\ref{ris2:0503yB2} shows the behaviour of the population as a function of the distribution of agents by opinion. It is observed that for initial values of the proportion of population agents holding opinion $m_1$, less than 0.5, the population reaches a stationary position where $x_N =(0.33; 0.67)$ (Figure~\ref{ris2:0503yB2}a)). On the other hand, for initial values of the proportion of population agents holding opinion $m_1$ of at least 0.5, the system reaches a stationary position where $x_N = (0.7; 0.3)$ (Figure~\ref{ris2:0503yB2}b)).

\section{Example 2. The Prisoner's Dilemma}

In the current experiment, the Prisoner's Dilemma in its economic interpretation is chosen as the base game describing the interaction of agents. In this game, the first strategy corresponds to the player's choice to cooperate and the second to the choice to defend.

Since the change in the state of the population as a function of the agents' opinions is considered in the model according to the formula~(\ref{An}), it is necessary to introduce two playoff matrices
\begin{equation}
A_0 =
    \begin{pmatrix}
3.5 & 1\\ 2 & 0.75
  \end{pmatrix},\,
  A_1 =
    \begin{pmatrix}
   4 & 1\\   4.5 & 1.25
  \end{pmatrix},
  \label{A0A1PrNumbers}
\end{equation}
for whose elements the relations~(\ref{param}) are true.
\begin{equation}
a_{11}^0 > a_{21}^0 ,\, a_{12}^0  > a_{22}^0,\,
a_{11}^1 < a_{21}^1,\, a_{12}^1 < a_{22}^1.
\label{param}
\end{equation}

The Prisoner's Dilemma game is characterized by the existence of a single Nash equilibrium state, and given the relations~(\ref{param}), for the game given by the matrix $A_1$($A_0$), this state is $(e^2,e^2)$($(e^1,e^1)$).

In the current numerical experiment, the parameters of the system~(\ref{syst}) take on values: 

\vskip-20pt
\begin{table}[H]
\caption{System parameters.\label{tab1}}
\newcolumntype{C}{>{\hsize=.27\hsize\centering\arraybackslash}X}
\newcolumntype{K}{>{\arraybackslash}X}
\small
\begin{tabularx}{\textwidth}{CCK}
\toprule
\textbf{Parameter}	& \textbf{Value}  & \textbf{Description}\\
\midrule
$\theta$   &2 & The rate of replenishment of resources by cooperates\\
$\psi$ & -1 & The rate at which defenders consume resources\\
$x_0$   &0.5 &Proportion of cooperators in the population at the initial time moment\\
$y_0$ & 0.6 & Proportion of agents with an opinion $m_1$ at initial time moment\\
$n_0$   &0.3  & Environmental condition at initial time moment \\
$B$ & $\left( \begin{smallmatrix}    0.5 & 0\\ 0 & 0.5\end{smallmatrix} \right)$ &  Agents with opinion $m_1$ ($m_2$) are assumed to trust only cooperators (defenders)\\
\bottomrule
\end{tabularx}
\end{table}
\unskip

\begin{figure}[h!]
\begin{minipage}[h]{0.49\linewidth}
\center{\includegraphics[width=1\linewidth]{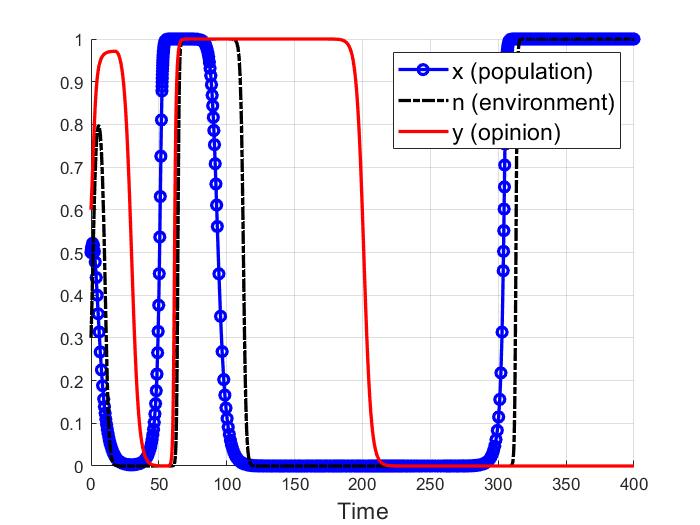} a) }
\end{minipage}
\hfill
\begin{minipage}[h]{0.49\linewidth}
\center{\includegraphics[width=1\linewidth]{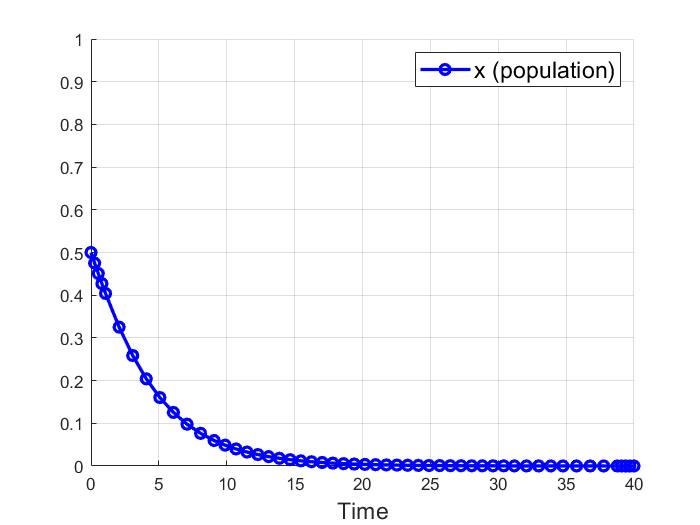} b) }
\end{minipage}
\caption{Illustration of the system state: a) with environment and distribution of agents between opinions; b) without influence of environment and opinions; $x(0) = 0.5$, $n=1$ is assumed.}
\label{pic1}
\end{figure}

As can be seen from the graph in Figure~\ref{pic1}a), at the initial time moment, the proportion of cooperating agents decreases to zero as the environment begins to enrich. However, as the proportion of defending agents in the population increases, the environmental resources decrease to zero. After several oscillations, the system reaches an equilibrium state $(e^1,e^1)$, all players have the opinion $m_2$, the environment is replenished, i.e. $n=1$.

Through a series of numerical experiments, it was found that the environment and the opinions of the agents have a significant impact on the stationary position of the population. In most cases, a change in the initial value of the population parameter, the environment, or the popularity of the opinions causes a change in the stationary position that the system reaches. The choice of the confidence matrix also plays an important role in the results of the simulation of changes in the population depending on the environment and the opinions of the agents.

\end{document}